\documentstyle[12pt]{article}
\newlength{\extralineskip}
\newcommand{\beq}{\begin{equation}}
\newcommand{\eeq}{\end{equation}}
\newcommand{\bd}{\begin{displaymath}}
\newcommand{\ed}{\end{displaymath}}
\def\bea{\begin{eqnarray}}
\def\eea{\end{eqnarray}}
\def\nn{\nonumber}
\def\ba{\beq\new\begin{array}{c}}
\def\ea{\end{array}\eeq}

\def\inbar{\,\vrule height1.5ex width.4pt depth0pt}
\def\IC{\relax\hbox{$\inbar\kern-.3em{\rm C}$}}
\def\IR{\relax{\rm I\kern-.18em R}}

\def\1{\relax{\rm 1\kern-.25em l}}

\def\Tr{{\rm Tr}}
\def\e{~{\rm e}}

\makeatletter
\newdimen\normalarrayskip              % skip between lines
\newdimen\minarrayskip                 % minimal skip between lines
\normalarrayskip\baselineskip
\minarrayskip\jot
\newif\ifold             \oldtrue            \def\new{\oldfalse}
\begin{document}
\begin{titlepage}
\thispagestyle{empty}
 
\begin{flushright}
PUPT-1982  
\end{flushright}

\vskip1cm
\begin{center}
{\LARGE \bf Exact Noncommutative KP and KdV Multi-solitons}\\
\vskip1cm
{\bf L.D. Paniak} \\
\bigskip
{\it Joseph Henry Laboratories, Princeton University \\
Princeton, New Jersey 08544, USA} \\
e-mail: paniak@feynman.princeton.edu
\vskip4cm

\begin{abstract}
We derive the Kadomtsev-Petviashvili (KP)
equation defined over a general associative algebra  
and construct its $N$-soliton solution. For the example of the Moyal 
algebra, we find multi-soliton solutions for arbitrary space-space 
noncommutativity. The noncommutativity of coordinates is shown to obstruct the 
general construction of a $\tau$ function for these solitons.
We investigate the two-soliton solution in detail and show that asymptotic 
observers of soliton scattering are unable to detect a finite spatial 
noncommutativity. An explicit example shows that a pair of solitons in 
a noncommutative background can be interpreted as several pairs of image 
solitons. Finally, a dimensional reduction gives the general $N$-soliton 
solution for the previously discussed noncommutative KdV equation.

\end{abstract}

\end{center}
\end{titlepage}
\newpage

\setcounter{page}1
\section{Introduction}

Noncommutativity of coordinates \cite{connes,cds} appears to be an important
ingredient of string theory.  For example, 
in Witten's formulation of open string field theory 
\cite{w1},  space-time noncommutativity leads to an elegant formulation of the 
string field equations of motion. More recently, 
the calculation \cite{gms} and analysis \cite{dasgupta,hklm}
of solitons in field theories simplified by strong noncommutativity has given 
evidence to support a conjecture of Sen \cite{sen} relating  
nonperturbative objects in open string field theory to (unstable) D-branes.
This success has fueled further developments on the subject which 
are reviewed in \cite{harvey}. 

Given the fundamental and calculational importance of noncommutative geometry, 
gaining a deeper understanding of its physical effects is essential.
Here we present a 
non-trivial, yet completely solvable model of soliton physics in 
a noncommutative geometry where the parameter controlling the noncommutativity
is left arbitrary. The solutions we find are not only of interest in 
answering questions in string theory but they also present a 
laboratory for detailed investigations into 
the impact the noncommutativity of coordinates has on soliton physics.

As our model we derive the analogue of the $2+1$-dimensional
Kadomtsev-Petviashvili (KP) equation
for soliton profiles which take values in a general associative algebra.  
The resulting noncommutative KP (ncKP) equation is an integrodifferential
equation. In spite of the complexity  of the ncKP equation, we are able to
explicitly construct an $N$-soliton solution via a modification of the 
trace method \cite{ow}. As a particular example we specify the Moyal
algebra of the two spatial coordinates to find the ncKP equation and
its associated $N$-soliton solution for noncommutative geometry.

We then proceed to examine the  one- and two-soliton solutions in detail.
The single soliton solution is unchanged in the deformation to 
noncommutative space due to the linear structure of the KP soliton.
In the general two-soliton solution, the noncommutative nature of the underlying
geometry leads to interesting effects. We show that the solution is 
periodic in a dimensionless parameter which controls the noncommutativity.
The asymptotic behaviour of soliton scattering is shown to be independent
of finite spatial noncommutativity, making it impossible to detect
a noncommutative geometry through the asymptotics. Finally, as an explicit
example we show how finite noncommutativity changes the physics of a pair
of solitons to resemble the interaction of four different pairs of solitons. 

Even though we can construct the $N$-soliton solution in an explicit
manner, space-space noncommutativity obstructs the direct construction 
of a $\tau$ (generating) function for these solutions.  
While it is possible that there is an analogue of
a $\tau$ function to generate soliton solutions and conserved currents for the  
noncommutative KP equation, its construction is more involved than in 
the commutative case. 
Consequently, our results suggest that the analysis of integrable systems in
noncommutative geometry is fundamentally more involved than in the 
commutative case.

While we concentrate on the ncKP equation defined in $2+1$ dimensions, 
it is a simple matter to eliminate a spatial direction and recover 
the noncommutative version of the Korteweg-de Vries (ncKdV) equation
\cite{dmh}. In this way, all of our calculations and observations
for the ncKP equation lead to analogous results for the ncKdV equation,
including its general $N$-soliton solution.
This is an efficient way to optimize our calculational efforts since
the ncKP and ncKdV equations with their respective solutions are
of similar complexity.

During the completion of this project, we discovered that the KP equation 
defined over a general finite-dimensional associative (matrix) algebra has 
been derived previously.
These and other results on noncommutative integrable systems may be found in 
the very recent book \cite{bak}.
 
\section{The noncommutative KP equation for general associative algebras}

\setcounter{equation}{0}

The (commutative) KP equation is a non-linear equation 
in two spatial coordinates which admits stable solitary wave (soliton)  
solutions. Theoretically it is interesting since 
it yields the simplest extension of the classic KdV
equation which is defined in  one space dimension. One way to 
construct the KP equation is through the following process.
Consider the differential equation
\beq
L \psi = \left( \frac{\partial^2}{\partial x^2} + u(x,y,t) \right) \psi = 
\lambda \psi
\label{ldefn}
\eeq
It is well-known that for reasonable choices of ``potential'' $u(x,y,t)$ and 
boundary conditions, there
is a discrete infinity of eigenvalues $\lambda$.  One can also view the 
time evolution of the wavefunction $\psi$
\beq
\partial_t \psi  = A \psi
\label{atflow}
\eeq
where we will choose  the operator $A$ to be linear and 
third order in derivatives of the coordinate $x$
\beq
A= \frac{\partial^3}{\partial x^3} + a_2(x,y,t) \frac{\partial^2}{\partial x^2} +
a_1(x,y,t) \frac{\partial}{\partial x} + a_0(x,y,t)
\label{adefn}
\eeq
 
The Lax method of generating the KP equation in particular, and more generally
other integrable non-linear equations, 
requires the mutual consistency of the two operators $L$ and $A$.
Considering the commutator of the  Lax pair, $L$ and $A$ and
requiring the eigenvalues to be constants of motion,  one finds Lax's equation 
\beq
[ L, A] = \partial_t  L -\partial_y A
\label{czs}
\eeq
In the present case, this equation allows one to solve for the coefficients
$a_i$ in (\ref{adefn}) and implies a consistency requirement
for the potential $u$. This constraint is known as the KP equation
\beq
0 = (u_t  +\frac{1}{4} u_{xxx} + \frac{3}{4} (u^2 )_x )_x + \frac{3}{4} u_{yy}
\label{kpu}
\eeq
Subsequently we will refer to  $u$ as the profile of the 
solitary wave (soliton) which solves (\ref{kpu}).

To obtain a coherent generalization of the KP equation for the case
where $u$ takes values in some general associative algebra,
we propose to deform the commutator in Lax's equation (\ref{czs}) to the 
$\star$-commutator.  This leads to a deformed version of Lax's equation
\beq
[ L , A ]_\star = L \star A  - A \star L= \partial_t L -\partial_y A
\label{nczs}
\eeq
We will leave the explicit definition of the $\star$-product arbitrary at
this point and only require it to commute with derivatives.

We follow as in the commutative case and 
substitute the definition of $L$ (\ref{ldefn}) and the ansatz 
(\ref{adefn}) for $A$
into the deformation of Lax's equation (\ref{nczs}).  
Again we arrive at a consistency requirement for the profile $u$ 
which in this case gives the noncommutative version of the KP equation (ncKP)
\beq
0 = (u_t  +\frac{1}{4} u_{xxx} + \frac{3}{4} (u \star u )_x + 
\frac{3}{4} [ u, \partial_x^{-1} u_y ]_\star )_x + \frac{3}{4} u_{yy}
\label{nckpu}
\eeq
While it can be readily checked that this equation reduces to the KP 
equation (\ref{kpu}) in the commutative limit, in general it is an 
integrodifferential
equation for the profile $u$. 
Note that due to the presence of the $\star$-commutator
term in (\ref{nckpu}), the noncommutative  KP equation 
is not the commutative version with $\star$-products naively replacing
commutative multiplication.  This `non-covariance' in moving to 
the noncommutative equation can only arise when one includes 
terms nonlinear in derivatives.  
Additionally, we remark that this equation is identical 
to the generalized KP one obtains when the profile $u$ is an element of
a generic matrix algebra \cite{dorfok,bak}.

We can obtain a 
differential equation  equivalent to (\ref{nckpu}) by moving to a potential formulation
of the problem. We do this by defining an auxiliary field $\phi$ where
\beq
u= 2 \phi_x
\eeq
Substituting into (\ref{nckpu}) transforms the ncKP equation into a
convenient form which we will use for subsequent calculations
\beq
0= (4 \phi_t + \phi_{xxx} + 6 \phi_x \star \phi_x )_x + 
6 [ \phi_x ,\phi_y ]_\star + 3 \phi_{yy}
\label{nckpphi}
\eeq
Once more, the $\star$-commutator term in this equation produces a contribution 
which is not present in the commutative case.

\section{Generating the $N$-soliton solution}
\setcounter{equation}{0}

There are many ways to solve the commutative KP equation but for our purposes the 
trace method \cite{ow} provides the most convenient starting point for 
generalizing to associative algebras.
The trace method assumes a formal power series for the $N$-soliton solution 
of the commutative KP equation in the potential formulation
\beq 
\phi = \sum_{n=1}^\infty \epsilon^n \phi^{(n)}
\label{phiansatz}
\eeq

In our case, substituting into (\ref{nckpphi}) and solving order by order in the 
formal expansion parameter $\epsilon$, one can verify the first few terms
\bea
\phi^{(1)} &=& \sum_{l=1}^N \phi_l \\
\phi^{(2)}& = & -\sum_{l_1,l_2 =1}^N  \frac{\phi_{l_1} \star \phi_{l_2}  }
{ (p_{l_1} + q_{l_2})}
\eea
where 
\beq
\phi_l = c_l \e^{ (p_l + q_l) x + (p_l^2 - q_l^2) y - (p_l^3 + q_l^3) t}
\label{recfac}
\eeq
Here we require $c_l (p_l + q_l)>0$, but otherwise $c_l$ is an arbitrary 
constant.

For the higher order terms, we have the recursive equation at $O(\epsilon^n)$
which gives an equation for $\phi^{(n)}$ in terms of lower order coefficients
\beq
[4 \partial_t + \partial_{xxx} + 3 \partial^{-1}_x \partial_{yy}] \phi^{(n)} =
-6 \sum_{r=1}^{n-1} \left( \partial^{-1}_x [ \phi_x^{(r)} ,\phi_y^{(n-r)} ]_\star +
\phi_x^{(r)} \star \phi_x^{(n-r)}  \right)
\eeq
A formal solution to this recursion relation, and the ncKP equation, 
is given by the coefficients
\beq
\phi^{(n)} =(-1)^{n-1} \sum_{l_1 \ldots l_n =1}^N 
\frac{ \phi_{l_1} \star \phi_{l_2}\star  \cdots \star \phi_{l_n}}
{ (p_{l_1} + q_{l_2}) (p_{l_2} + q_{l_3})  \cdots (p_{l_{n-1}} + q_{l_n}) }
\label{phinstar}
\eeq

This claim can be proved through substitution and by using the following 
algebraic identity
\bea
\lefteqn{(p_1 + q_1 + \cdots +p_n + q_n )^4 - 4  (p_1 + q_1 + \cdots +p_n + q_n )
(p_1^3 + q_1^3 + \cdots +p_n^3 + q_n^3 ) } && \nn \\
&& +  3 (p_1^2 - q_1^2 + \cdots +p_n^2 - q_n^2 )^2  \\
& = & 6 \sum_{r=1}^{n-1} (p_r + q_{r+1} ) [ (p_1 + q_1 + \cdots + p_r + q_r )
(p_{r+1}^2 - q_{r+1}^2 + \cdots +p_n^2 - q_n^2 ) 
\nn \\
&& - (p_{r+1} + q_{r+1} + \cdots +p_n + q_n )(p_{1}^2 - q_{1}^2 + \cdots +p_r^2 - q_r^2 ) 
\nn \\
&& + (p_1 + q_1 + \cdots +p_r + q_r )
(p_{r+1} + q_{r+1} + \cdots +p_n + q_n )(p_1 + q_1 + \cdots +p_n + q_n )]
\nn
\eea
The identity itself can be proved through a  process of
matching the coefficients on either side of the equality.
This procedure is simplified by noticing that each side is a quartic 
polynomial in the variables $p_k$ and $q_k$.  
In the limit of a commuting algebra, we can use the symmetry of the summand in 
(\ref{phinstar}) to 
recover the known result for the commutative case \cite{ir,ow}.
 
It is interesting to observe that even though the noncommutative version
of the KP equation (\ref{nckpphi}) 
involves certain `non-covariant' terms (the $\star$-commutator), 
its solution (\ref{phinstar})
is just the commutative solution with the $\star$-product replacing
the usual commutative product. 

\section{The KP equation in noncommutative geometry}
\setcounter{equation}{0}

We will now specify the associative product of the previous section to be the 
Moyal product for two noncommuting spatial coordinates.
In particular, the $\star$-product of two functions is explicitly defined in terms of the 
non-commutativity parameter $\theta$ by
\beq
f(x,y) \star g(x,y) \equiv \left. \exp{\left[ \frac{ \theta}{2} 
( \partial_x \partial_{y^\prime} - \partial_{y} \partial_{x^\prime} ) \right]}
f(x,y) g(x^\prime, y^\prime) \right|_{x=x^\prime, y= y^\prime}
\eeq

For many calculations the parameterization in terms of $p_l$ and $q_l$ as introduced
in (\ref{recfac}) will 
be most convenient.  These parameters can be used to construct physically relevant 
quantities such as the wavevector $\vec{k} = (k_x, k_y)$ and the frequency $\omega$ of the $l^{\rm th}$ soliton
through
\bea
(k_x)_l & = & p_l + q_l \\
(k_y)_l & = & p_l^2 - q_l^2 \\
\omega_l & = & p_l^3 + q_l^3
\eea
With these definitions we can now introduce the cross-product notation for 
soliton wavevectors, which will simplify calculations
\bea
\vec{k}_{n} \times  \vec{k}_{m} &=& (k_x)_n (k_y)_m - (k_x)_m (k_y)_n \\
&=& (p_n + q_n)(p^2_m - q^2_m) - (p^2_n - q^2_n) (p_m + q_m) \nn
\eea

Given these definitions, the $N$-soliton solution of the ncKP equation for the case 
of noncommuting spatial coordinates can be calculated from (\ref{phiansatz}) and
(\ref{phinstar})
\beq
\phi_N = \sum_{n=1}^\infty (-1)^{n-1} \sum_{l_1 \ldots l_n =1}^N 
\frac{ \e^{\frac{\theta}{2} \sum_{i<j} \vec{k}_{l_i} \times  \vec{k}_{l_j} }
\phi_{l_1} \phi_{l_2}  \cdots \phi_{l_n}}
{ (p_{l_1} + q_{l_2}) (p_{l_2} + q_{l_3})  \cdots (p_{l_{n-1}} + q_{l_n}) }
\label{phinphase}
\eeq

The result of noncommutativity in the spatial coordinates is the appearance
of an index-dependent factor 
\beq
\e^{\frac{\theta}{2} \sum_{i<j} \vec{k}_{l_i} \times  \vec{k}_{l_j}}
\label{ncphase}
\eeq
which is familiar from the calculation of Feynman diagrams in noncommutative 
field theories \cite{filk,iikk,MvRS} or open string amplitudes in a background
magnetic field \cite{sw}.
This noncommutative factor has important implications for the solution. 
Following as in the commutative case, we can better see the structure of the 
solution by introducing the matrix
\beq
B_{lm} = \frac{c_m  \e^{p_l x} \e^{ q_m x +(p_m^2 -q_m^2) y - (p_m^3 +q_m^3) t} }
{p_l + q_m}
\label{bdefn}
\eeq
where $c_m (p_m + q_m)>0$ as before.
With this definition we can rewrite the $N$-soliton solution (\ref{phinphase})
in a concise manner
\beq
\phi_N = \sum_{n=1}^\infty (-1)^{n-1}
\sum_{l_1 \ldots l_n =1}^N 
\e^{\frac{\theta}{2} \sum_{i<j} \vec{k}_{l_i} \times  \vec{k}_{l_j} }
B_{l_1 l_2} B_{l_2 l_3} \cdots B_{l_{n-1} l_n} \frac{\partial}{\partial x} 
B_{l_n l_1}
\label{bsoln}
\eeq

In the commutative case $(\theta=0)$,
the sum over indices $l_i$ is a trace over the product of $B$ factors.  
Moreover, the 
derivative can be extracted from the sum and the series over $n$ summed
\beq
\phi_N = \sum_{n=1}^\infty (-1)^{n-1}
{\rm Tr} \left(  B^{n-1} \frac{\partial}{\partial x} B \right)
= \frac{\partial}{\partial x} {\rm Tr} \ln{ (\1 + B)} 
= \frac{\partial}{\partial x}   \ln{ \det{(\1 + B) }}
\eeq
Consequently, we obtain a well-known result: the $N$-soliton 
profile for the KP equation can be recovered from logarithmic derivatives 
of a generating function
\beq
u = 2 \frac{\partial^2}{\partial x^2} \ln{ \det{( \1+ B)} }
\label{solsum}
\eeq
Customarily, the generating function is labeled $\tau$  
\beq
\tau =  \det{( \1+ B)} 
\eeq
and it plays a fundamental role in the theory of integrable systems 
(see \cite{newell} for example).  

In the present noncommutative case, the appearance of  noncommutative factors in 
(\ref{bsoln}) leads to substantial changes in the character of the matrix 
series.  If one considers each term  in (\ref{bsoln}) as a matrix chain with 
periodic boundary conditions, then noncommutativity can be seen to introduce
non-local, anisotropic interactions between the sites $(l_i)$ on 
the chain. While long-range interactions complicate the evaluation of 
(\ref{bsoln}), it is the anisotropy of the interactions which has the 
most profound effect. For any non-zero $\theta$,  each factor of $B$ in the 
series is distinguished, making it impossible to extract the derivative as
in the commutative case.
This makes the direct identification of a $\tau$ function for the 
noncommutative KP equation impossible in the current framework. 
It would appear that
there are significant changes to the theory of integrable systems as soon as 
spatial noncommutativity is turned on. This same type of obstruction was observed 
previously in the noncommutative KdV equation \cite{bak}.

\section{One Soliton}
\setcounter{equation}{0}

In the case of $N=1$, the single soliton, there is only one wavevector, $\vec{k}$
and hence the factor (\ref{ncphase}) associated with the spatial
noncommutativity vanishes.  Consequently, the 
noncommutative one-soliton solution coincides with the solution in 
commutative space. From equation (\ref{solsum}) we find the well-known
one-soliton profile
\beq 
u_1(x,y,t) = \frac{1}{2} k_x^2 
{\rm sech}^2 \left(\frac{1}{2} (k_x x + k_y y - \omega t) + \delta \right)
\label{1soliton} 
\eeq
This is not surprising given the nature of the KP soliton.  It is a one-dimensional
object which moves in the $x-y$ plane with constant wavevector $\vec{k}$.
Consequently, the profile (\ref{1soliton}) depends only on a single argument.
By changing coordinates to those parallel and perpendicular to the soliton, it is
easy to show that $\star$-products involving  $u_1$ reduce to commutative
multiplication.  Analogous behaviour is seen for noncommutative versions of the 
KdV \cite{bak,dmh} and nonlinear Schr\"odinger equations \cite{dmhnls}.

This argument may be extended to cases with an arbitrary number of parallel
solitons, where the cross-products of all wavevectors vanish 
\beq
\vec{k}_{n} \times  \vec{k}_{m} =0 ~~\forall~n,m
\eeq
With this condition satisfied, from (\ref{bsoln}) it is clear that the solution for 
parallel solitons will reduce to that of the commutative 
case. Note that even though the effect of spatial noncommutativity is trivial 
in this case, the dynamics of parallel solitons in the commutative case is 
non-trivial.

\section{Two Solitons}
\setcounter{equation}{0}

The two-soliton solution of the ncKP equation 
is much more interesting since the noncommutativity 
generally affects the physics of the colliding solitons. This minimally complicated 
case has only one new factor due to noncommutativity which involves the 
wavevectors of the two solitons
\beq
\e^{\frac{\theta}{2}  \vec{k}_{1} \times  \vec{k}_{2} }
\eeq
The labeling of the solitons is made unique by requiring the cross-product of the 
wavevectors to be non-negative
\beq
\vec{k}_{1} \times  \vec{k}_{2} = (k_x)_1 (k_y)_2 - (k_y)_1 (k_x)_2 \geq 0
\eeq
For further computation it is useful to introduce a 
dimensionless noncommutativity parameter $\Theta$  
\beq
\Theta = \frac{\theta }{2} {\vec{k}_{1} \times  \vec{k}_{2}}
\eeq
As noted above, the effective noncommutativity vanishes for non-zero $\theta$ 
when the wavevectors of the solitons are parallel: 
$\vec{k}_{1} \times  \vec{k}_{2} =0$.  
 
In the case of the two-soliton solution, there is a substantial simplification
of the structure of the nonlocality in the index space of the solution as 
compared to the general noncommutative $N$-soliton solution (\ref{bsoln}).
For two solitons the nonlocality can be re-interpreted purely as anisotropy in the 
index space.  More precisely, since $l_i = 1,2$ we have from the noncommutative 
factor (\ref{ncphase})
\beq
\frac{\theta}{2} \sum_{i<j}^n \vec{k}_{l_i} \times  \vec{k}_{l_j}
= \Theta  \sum_{i<j}^n 
\frac{ \vec{k}_{l_i} \times  \vec{k}_{l_j}}{\vec{k}_1 \times  \vec{k}_2}
=  \Theta \sum_{i<j}^n (l_j - l_i )   
=  \Theta \sum_{j=1}^n \sum_{i=1}^j (l_j - l_i )   
= \Theta \sum_{j=1}^n (2 j - n -1) l_j  
\label{2solident} 
\eeq

The advantage gained by this observation is that it allows for 
the two-soliton solution to be written as the trace over a product of matrices.  
Using (\ref{bsoln}) and (\ref{2solident})  we find
\bea
\phi_2 & = & \sum_{n=1}^\infty (-1)^{n-1} \! \sum_{l_1 \ldots l_n =1}^2 \!
\e^{ -\Theta (n-1) l_1}  B_{l_1 l_2} \e^{ -\Theta (n-3) l_2} B_{l_2 l_3} \cdots 
\e^{  \Theta (n-3) l_{n-1}} B_{l_{n-1} l_n} 
\e^{  \Theta (n-1) l_{n}} \frac{\partial}{\partial x} B_{l_n l_1} \nn \\
& = & \sum_{n=1}^\infty 
(-1)^{n-1} {\rm Tr} \left( \e^{\frac{\Theta}{2} ( n -1) {\bf \sigma_3}} B 
\e^{\frac{\Theta}{2} ( n -3) {\bf \sigma_3}} B \cdots
\e^{-\frac{\Theta}{2} ( n -3) {\bf \sigma_3}} B 
\e^{-\frac{\Theta}{2} ( n -1) {\bf \sigma_3}} \frac{ \partial B}{\partial x} 
\right)
\label{2bsoln}
\eea
where the matrices $B$ are defined in (\ref{bdefn}).

In this form we can already see some of the analytic structure of 
the two-soliton solution as a function of the effective noncommutativity.
Consider the effect of an imaginary shift in $\Theta$
\beq
\Theta \rightarrow \Theta + 2 \pi i m
\eeq
where $m$ is some integer. From the form (\ref{2bsoln}) we see that  
each of the factors associated with noncommutativity is changed by a factor
\beq
\e^{\frac{\Theta}{2} (n + 1-2 k) \sigma_3} \rightarrow 
(-1)^{m(n+ 1 - 2 k)} \e^{\frac{\Theta}{2} (n + 1-2 k) \sigma_3}
\eeq
It is straightforward to show that the factor 
$(-1)^{m(n+ 1 - 2 k)}$ has no effect on the solution (\ref{2bsoln}) as
it cancels out in each term of the series. Furthermore, it is 
not difficult to convince oneself that (\ref{2bsoln}) is not periodic 
under shifts in $\Theta$ of less than $2 \pi$. 
Hence 
the two-soliton solution of the ncKP equation 
is a periodic function of $\Theta$ with period $2\pi$.
This is a simple prototype of a multi-dimensional periodicity 
which arises in the $N$-soliton solution (\ref{bsoln}).

In the limit of $\Theta$ vanishing, the derivative can be extracted and the 
series (\ref{2bsoln}) 
summed leading to the appearance of the $\tau$ function, as seen 
previously.  In addition, it is straightforward to calculate the 
first correction to the two-soliton solution for non-vanishing 
$\Theta$. Introducing the  phases of each of the two solitons
\bea
\eta_1 &=& \frac{1}{2} \left( 
(k_x)_1 x + (k_y)_1 y - \omega_1 t + \ln{\left( \frac{ c_1}{p_1+q_1} \right)}
\right)   \\
\eta_2 &=&  \frac{1}{2} \left(  
(k_x)_2 x + (k_y)_2 y - \omega_2 t + \ln{\left( \frac{ c_2}{p_2+q_2} \right)}
\right)  
\eea
and defining the real quantity $\Delta$ by
\beq
\e^{-2 \Delta} = 1 - \frac{ (p_1+q_1) (p_2 + q_2) }{ ( p_1 + q_2) (p_2 + q_1)}
=\frac{ (p_1-p_2) (q_1 - q_2) }{ ( p_1 + q_2) (p_2 + q_1)} \geq 0
\eeq
we can express the leading terms of a perturbative expansion in the
dimensionless noncommutativity parameter  
\bea
u_2  &=& 2 \frac{\partial^2}{\partial x^2} \ln{( 1 + \e^{2 \eta_1} +
\e^{2 \eta_2} +\e^{2 (\eta_1+ \eta_2   - \Delta) }) }
\label{phiexp} \\
&& + 2 \frac{\partial }{\partial x } \left( \Theta ~\frac{( p_1 + q_2 - p_2 -q_1) 
( 1 - \e^{-2 \Delta}) \e^{2 (\eta_1 + \eta_2)} }{(1 + \e^{2 \eta_1} + 
\e^{2 \eta_2} + \e^{2 (\eta_1 +  \eta_2  - \Delta)})^2 } \right)
+ O(\Theta^2) \nn
\eea
 
This calculation tells us several things.  To first order, the noncommutative
two-soliton solution is a smooth deformation of the commutative solution
and reduces to it in the limit of vanishing $\Theta$. 
It is worth noting the structure of the 
first correction in $\Theta$. It is antisymmetric under exchange of soliton
label and it vanishes for $\Delta=0$. The antisymmetry is a natural effect
of the manner in which the spatial noncommutativity  arises in the solution.  
The vanishing of the correction with $\Delta$ is true to all orders in $\Theta$ 
as we shall soon see. 

In order to perform further calculations in the noncommutative case, it is 
convenient to conjugate the matrices in each term of the series 
(\ref{2bsoln}) to obtain the form
\beq
\phi_2=\sum_{n=1}^\infty (-1)^{n-1} {\rm Tr} \left( \prod_{k=1}^{n-1} C_{k,n} D
\right)
\label{phidiag}
\eeq
Here we have defined the matrices $C_{k,n}$ and $D$ by
\beq
C_{k,n} =  \left( \begin{array}{cc} 
\e^{  2 \eta_1 } & \sqrt{ 1 - \e^{-2 \Delta} } \e^{  \eta_1 + \eta_2 +  
\Theta k( n-k)    } \\
\sqrt{ 1 - \e^{-2 \Delta} } \e^{ \eta_1 + \eta_2 -
\Theta k( n-k) } & \e^{  2 \eta_2 } 
\end{array} \right)
\eeq
and
\beq
D  =  \left( \begin{array}{cc} 
(p_1+q_1) \e^{  2 \eta_1 } & (p_1+q_2) 
\sqrt{ 1 - \e^{-2 \Delta} } \e^{  (\eta_1 + \eta_2)  } \\
(p_2 + q_1) \sqrt{ 1 - \e^{-2 \Delta} } \e^{  (\eta_1 + \eta_2)   } & 
(p_2 +q_2) \e^{ 2 \eta_2 } 
\end{array} \right)
\eeq

Generally speaking, (\ref{phidiag}) is a deformation of the geometric
series by spatial noncommutativity. Summing this matrix series exactly 
appears to be a difficult task. There are however several important 
observations to be made about noncommutative soliton physics from 
the form of the matrices $C_{k,n}$. 

First of all, we see that the effect of the noncommutative factors
lies solely in the off-diagonal components of $C_{k,n}$ and that these
components vanish for $\Delta =0$.  The limit of vanishing $\Delta$ is 
known from the commutative case to lead to the reduction of the 
two-soliton problem to that of two, non-interacting single solitons. In this 
way we see another manifestation of the observation made for parallel 
solitons: spatial noncommutativity is irrelevant without interaction.

Physically, the condition $\Delta=0$ can be realized in several ways.
First, if the x-component of the wavevector for one, or both of
the solitons is zero  
\beq
(k_x)_1 = p_1 + q_1 =0 ~~~ \mbox{and, or}~~~(k_x)_2 = p_2 + q_2 =0
\eeq
Since $k_x$ also gives the amplitude of each of the single solitons, we
see that in this limit one, or both of the solitons disappear. In addition
to this somewhat degenerate case, $\Delta=0$ if 
\beq
p_1 + q_2 \rightarrow \infty ~~~ \mbox{and, or}~~~ p_2 + q_1 \rightarrow \infty
\eeq
In the first case, the first soliton is made infinitely 
narrow \cite{ow}.  Likewise for the second case and the second soliton.

\subsection{Soliton scattering in noncommutative space}

In the commutative case, the interaction of two KP solitons takes place where the 
one-dimensional solitons cross in the $x$-$y$ plane. 
Far away from this interaction
region, the solitons are isolated from each other and we expect that locally, 
solutions are essentially the 
one-soliton solution.  In the commutative case, direct calculation \cite{ow,ir}
shows this to be the case, up to an important shift.  The asymptotic two-soliton
profile is
\beq
u_2(x,y,t) = 
\left\{ \begin{array}{c} \frac{(k_x)_1^2}{2} {\rm sech}^2  \eta_1   + 
\frac{(k_x)_2^2}{2} {\rm sech}^2( \eta_2 - \Delta) ~~~\mbox{as} ~~y  
\rightarrow \infty \\
\frac{(k_x)_1^2}{2} {\rm sech}^2( \eta_1 - \Delta ) + 
\frac{(k_x)_2^2}{2} {\rm sech}^2 \eta_2  ~~~\mbox{as}~~ y  \rightarrow -\infty
\end{array} \right.
\label{comasymp}
\eeq
The interpretation of this result reveals that the size of the region where the
two solitons interact is proportional to $\Delta$. This produces an offset
of the same size between pre- and post-interacting branches of each soliton,
as seen in the asymptotic solutions. 

In the case of noncommutative geometry, we have already seen that the one-soliton
solution coincides with the commutative solution so that the asymptotic solution
(\ref{comasymp}) is possible. Here we show indeed that (\ref{comasymp}) gives
the asymptotic behaviour in the noncommutative geometry.

The matrix $C_{k,n}$ which appears in the solution (\ref{phidiag}) can
be diagonalized since for imaginary $\Theta$ it is Hermitian. 
\beq
C_{k,n}=  \Omega_{k,n}
\left( \begin{array}{cc} \lambda_+ & 0 \\ 0 &\lambda_- \end{array} \right)
\Omega^{-1}_{k,n}
\label{cdiag} 
\eeq
Here the eigenvalues, $\lambda_\pm$ are independent of the indices $k$ and 
$n$
\beq
\lambda_\pm = \frac{1}{2} \left[ \e^{2 \eta_1} + \e^{2 \eta_2} \pm
\sqrt{ ( \e^{2 \eta_1} + \e^{2 \eta_2})^2 - 4 \e^{ 2( \eta_1+ \eta_2 - 
\Delta) } } \right]
\eeq
The unit-determinant matrix $\Omega_{k,n}$ 
\beq
\Omega_{k,n} =\left( \begin{array}{cc} \cos{\psi} & \e^{\Theta k (n-k)} 
\sin{\psi} \\ -\e^{-\Theta k (n-k)} \sin{\psi} &\cos{\psi}\end{array} \right)
\eeq
depends on the angle $\psi$   defined by
\beq
\cos{ 2 \psi} = \sqrt{ \frac{ ( \e^{2 \eta_1} -\e^{2 \eta_2})^2}
{ ( \e^{2 \eta_1} + \e^{2 \eta_2})^2 - 4 \e^{ 2( \eta_1+ \eta_2 - 
\Delta) } } }
\label{psidefn}
\eeq

In the commutative case, this diagonalization is independent of 
$k$ and $n$ and the solution depends only on the eigenvalues 
$\lambda_\pm$.  This is not true in general in the noncommutative 
case and we see that the noncommutative soliton has more degrees of 
freedom. In the limits of $\eta_i \rightarrow \pm \infty$, which are 
correlated with the asymptotic limits of $y \rightarrow \pm \infty$ \cite{ir},  
we see a significant simplification from (\ref{psidefn}) as 
$\cos{2 \psi} \rightarrow 1$. Consequently, to leading order,
(\ref{cdiag}) is independent of $\Theta$ and so the evaluation of 
the solution in (\ref{phidiag}) agrees with the commutative case and
gives the asymptotic behaviour (\ref{comasymp}).
 
Since the asymptotic form of the two-soliton solution is unchanged
for finite spatial noncommutativity, it is impossible for observers
far from the interaction region to use soliton scattering to detect
spatial noncommutativity.  Even though the interaction of the two 
solitons can be detected asymptotically due to the shift $\Delta$, 
the fact that this interaction took place in a noncommutative background 
can not. While the effects of noncommutative geometry are not
visible at long distances, they are certainly not weak, as we shall
see in an explicit solution in the next section.

\subsection{Closed form solutions for rational imaginary $\Theta$}

An extension of the logic we used to show that the two-soliton solution 
for the ncKP  equation is periodic in imaginary $\Theta$ 
can also lead us to closed form exact solutions when $\Theta$ is a 
rational multiple of $i \pi$
\beq
\Theta = i \pi \frac{p}{q} 
\eeq
where $p$ and $q$ are relatively prime integers.
This simplification can be seen by considering the general noncommutative 
factor from the solution (\ref{2bsoln}). In this case
\beq
\e^{\frac{i \pi p}{2 q} (n + 1-2 k) \sigma_3} = 
\cos{ \left(\frac{\pi p}{2 q} (n + 1-2 k) \right) } \1 + 
i \sin{ \left(\frac{\pi p}{2 q} (n + 1-2 k) \right) } \sigma_3
\label{trigident}
\eeq
The presence of rational multiples  of $\pi$ here guarantees that there
is only a finite number of unique factors which arise in the evaluation
of (\ref{2bsoln}). Consequently, the infinite series in (\ref{2bsoln})
can be split into distinct sums over $2 q$ equivalence classes.  
Each of these sums can be evaluated in terms of 
elementary functions.  The simplest example of this is when $p=q=1$
and the identity (\ref{trigident})
allows one to write (\ref{2bsoln}) as the sum of contributions
from odd $n$ and even $n$
\beq
\phi_2(\Theta=i \pi) = \frac{\partial}{\partial x}
\left[ \sum_{n ~{\rm odd}} \frac{(-1)^{n-1}}{n} \Tr B^n +
\sum_{n ~{\rm even}} \frac{(-1)^{n-1}}{n} \Tr ( \sigma_3 B)^n  \right]
\eeq
The series which appear can be summed exactly in terms of logarithms.
Using the representation of the components of $B$ in terms of $\eta_1$, 
$\eta_2$ and $\Delta$, we have the explicit form for the two-soliton
profile for this case
\bea
\lefteqn{u_2( \Theta = i \pi)} &&  \label{pisoln} \\
&=& \frac{\partial^2}{\partial x^2} \left(
\ln{  ( 1 + \e^{2 \eta_1} +
\e^{2 \eta_2} +\e^{2 (\eta_1+ \eta_2   - \Delta) })}
- \ln{( 1 - \e^{2 \eta_1} -
\e^{2 \eta_2} +\e^{2 (\eta_1+ \eta_2   - \Delta) })  } \right.
\nn \\
& & \left. +\ln{( 1 + \e^{2 \eta_1} -
\e^{2 \eta_2} -\e^{2 (\eta_1+ \eta_2   - \Delta) }) }
+\ln{( 1 - \e^{2 \eta_1} +
\e^{2 \eta_2} -\e^{2 (\eta_1+ \eta_2   - \Delta) })} \right)
\nn
\eea

While the solution (\ref{pisoln}) contains
singularities in the x-y plane and hence is not physically acceptable, it
does give us a glimpse of the effect of finite spatial noncommutativity
in an exactly solvable model.  With the form of the commutative solution 
(Eqn. \ref{phiexp} with $\Theta=0$) 
in mind, we see that (\ref{pisoln}) describes not a single
pair of solitons, but rather four pairs of solitons; one pair for each 
logarithm. Also it is a simple matter to check that the asymptotic 
$(\eta_i \rightarrow \pm \infty)$ behaviour
of the solution agrees with that of the commutative case, as expected 
from the general arguments above. As well we note that (\ref{pisoln})
is symmetric under the exchange of soliton label; a feature which is 
only present for $\Theta =0$ and $\pi$.

We see explicitly that the two-soliton  solution for $\Theta = i \pi$
can be written in terms of a generating $\tau$ function
\bea
\tau^2 
&=&  ( 1 + \e^{2 \eta_1} +
\e^{2 \eta_2} +\e^{2 (\eta_1+ \eta_2   - \Delta) })^2  \\
&& + 
4 ( 1- \e^{-2 \Delta} ) e^{ 2(\eta_1 + \eta_2)} \frac{ 
1 + \e^{2 \eta_1} +\e^{2 \eta_2} +\e^{2 (\eta_1+ \eta_2   - \Delta) } }
{1 - \e^{2 \eta_1} -
\e^{2 \eta_2} +\e^{2 (\eta_1+ \eta_2   - \Delta) } } 
\nn
\eea
In this form it is clear that with $\Delta=0$ the solution reduces
to the commutative case.
This example shows that it is possible to have a $\tau$-function for 
non-vanishing spatial noncommutativity but we expect this to be a unique 
feature for $\Theta = i \pi$ and directly tied to the symmetry of the 
solution under the exchange of soliton labels. 

\section{Reduction to noncommutative KdV}
\setcounter{equation}{0}

As a byproduct of our solution of the ncKP equation, we can find the N-soliton
solution for the previously discussed \cite{dmh,legare} noncommutative KdV equation 
(ncKdV). In the commutative 
case one can recover the KdV equation from the KP equation through a simple
dimensional reduction.  In the present case, 
assuming a soliton profile independent of the $y$-coordinate, $u(x,t)$, the ncKP 
equation (\ref{nckpu}) can be seen to reduce to the ncKdV equation as found in 
\cite{dmh}
\beq
0 = u_t  +\frac{1}{4} u_{xxx} + \frac{3}{4} (u \star u )_x  
\label{nckdveqn}
\eeq
At this point we have not yet defined the $\star$-product and this equation 
holds for any associative product.  In order to obtain the KdV equation in 
noncommutative geometry we specify the space-time $\star$-product
\beq
f(x,t) \star g(x,t) \equiv \left. \exp{\left[ \frac{ \theta}{2} 
( \partial_x \partial_{t^\prime} - \partial_{t} \partial_{x^\prime} ) \right]}
f(x,t) g(x^\prime, t^\prime) \right|_{x=x^\prime, t= t^\prime}
\eeq

We can easily produce an $N$-soliton solution to (\ref{nckdveqn}) 
in noncommutative space-time 
by restricting the $y$-component of each soliton wavevector to vanish in the
ncKP $N$-soliton solution (\ref{phinphase})  
\beq
(k_y)_l  = p_l^2 - q_l^2 =0
\eeq
Choosing $p_l=q_l$, we can eliminate all $y$-dependence and obtain a solution 
which only depends on $x$ and $t$
\beq
u(x,y) = 2 \frac{\partial}{\partial x} \left( \sum_{n=1}^\infty 
(-1)^{n-1} \sum_{l_1 \ldots l_n =1}^N 
\e^{ \frac{\theta}{2} \sum_{i<j}  {\vec k}_{l_i} \times {\vec k}_{l_j}}
\frac{ \chi_{l_1}   \chi_{l_2}  \cdots  \chi_{l_n}}
{ (p_{l_1} + p_{l_2}) (p_{l_2} + p_{l_3})  \cdots (p_{l_{n-1}} + p_{l_n}) }
\right) 
\label{nckdvsoln}
\eeq
with 
\beq
\chi_l = c_l \e^{ 2 p_l  x   - 2 p_l^3 t}
\eeq
and
\beq
{\vec k}_l = (2 p_l, - 2 p_l^3 )
\eeq
An analogous solution appeared in \cite{bak} in the context of the matrix 
KdV equation.

From the explicit solution (\ref{nckdvsoln}) of the ncKdV equation we can 
immediately recover an important result of \cite{dmh}: the solution is a
symmetric function of the noncommutativity parameter $\theta$.  Here we 
see that a change in sign of $\theta$ can be absorbed in the exponential 
factor
\beq 
\e^{ \frac{\theta}{2} \sum_{i<j}  {\vec k}_{l_i} \times {\vec k}_{l_j}}
\eeq
by reversing the order of the summation indices $l_j$
\beq
l_1, l_2, \ldots, l_n ~~~\rightarrow ~~~ l_n, l_{n-1}, \ldots, l_1
\eeq
By inspection, the remainder of the summand in (\ref{nckdvsoln}) is invariant 
under such a reordering. Hence the $N$-soliton solution of the ncKdV equation
is an even function of $\theta$.  
Note that this property is not shared by the ncKP equation discussed 
previously.

\section{Future directions}
\setcounter{equation}{0}

While we have not been able to find a closed form  
for the matrix series (\ref{2bsoln}) which gives the general two-soliton
solution, we expect progress is possible.  Given the symmetry of the series,
it should be possible to at least obtain an equivalent one-dimensional series
representation.
It would be desirable to have a more tractable 
form  of the solution in order to answer a number of interesting questions.  

First of all, having a closed form would 
certainly answer the question of whether the formal solution we have found 
is well-defined for non-zero $\Theta$. While we have seen perturbatively
and for $\Theta = i \pi$ that the two-soliton solution is reasonably 
well-behaved, it is not clear whether this is true in general.
Furthermore, 
with a better understanding of the two-soliton solution for finite $\Theta$,
we would be in a position to investigate the limit of large noncommutativity.
Finding this limit would allow one to make comparisons to recent 
multi-soliton solutions \cite{multisol1,multisol2,multisol3,multisol4} 
found at infinite noncommutativity.

A closed form for the two-soliton solution (\ref{2bsoln}) would also help 
to answer the question of whether a $\tau$ function can be found for the ncKP
solitons.  We are in the unusual position of having the $N$-soliton
solution for a non-linear equation without even knowing 
if a $\tau$-function exists. Clearly, if a generating function 
exists it is not simply calculable using the methods we have used here.
Presumably, knowledge of the $\tau$-function would allow for a complete
understanding of the integrable structure of the ncKP (and ncKdV) equation
as well as the calculation of conserved quantities.

An interesting application of our results would be to further investigate
non-perturbative objects in open $N=2$ string theory with a $B$-field 
\cite{lps1,lps2}.
Using the relationship between this string theory and $2+2$-dimensional 
noncommutative self-dual Yang-Mills theory \cite{lps1}, one can dimensionally 
reduce the gauge theory to recover \cite{legare} a matrix generalization of 
the noncommutative KdV equation we have solved here. As shown 
in \cite{lps2}, soliton solutions in the lower dimensional theory correspond
in the string theory to D-branes with a background magnetic field restricted
to their world-volume.

\section*{Acknowledgments}
We thank A.W. Peet for her comments and hospitality at the University of Toronto 
where part of this work was completed. We have benefited from numerous 
discussions with V.N. Muthukumar and comments by P. Charland, 
F. Ferrari, O. Lechtenfeld, J. Liu and E. Verlinde.  
This work was supported in part by the US Department of Energy grant  
DE-FG02-91ER40671.

%\newpage 

\end{document}